\documentclass[openacc]{rstransa}
\usepackage[numbers,sort&compress]{natbib}
\usepackage{microtype}
\usepackage{graphicx}
\usepackage{amsmath}
\graphicspath{ {./figures/} }
\usepackage{float}
\usepackage{lmodern}

\begin{document}

\title{A framework for probabilistic weather forecast post-processing across models and lead times using machine learning}

\author{
Charlie Kirkwood$^{1}$, Theo Economou$^{1}$, Henry Odbert$^{2}$ and Nicolas Pugeault$^{3}$}

\address{$^{1}$College of Engineering, Mathematics and Physical Sciences,
	University of Exeter, Exeter, UK \\
$^{2}$Met Office, Exeter, UK\\
$^{3}$School of Computing Science, University of Glasgow, Glasgow, UK}

\subject{meteorology, statistics, machine learning}


\corres{Charlie Kirkwood\\
\email{c.kirkwood@exeter.ac.uk}}

\begin{abstract}
Forecasting the weather is an increasingly data intensive exercise. Numerical Weather Prediction (NWP) models are becoming more complex, with higher resolutions, and there are increasing numbers of different models in operation. While the forecasting skill of NWP models continues to improve, the number and complexity of these models poses a new challenge for the operational meteorologist: how should the information from all available models, each with their own unique biases and limitations, be combined in order to provide stakeholders with well-calibrated probabilistic forecasts to use in decision making?

In this paper, we use a road surface temperature example to demonstrate a three-stage framework that uses machine learning to bridge the gap between sets of separate forecasts from NWP models and the `ideal' forecast for decision support: probabilities of future weather outcomes. First, we use Quantile Regression Forests to learn the error profile of each numerical model, and use these to apply empirically-derived probability distributions to forecasts. Second, we combine these probabilistic forecasts using quantile averaging. Third, we interpolate between the aggregate quantiles in order to generate a full predictive distribution, which we demonstrate has properties suitable for decision support. Our results suggest that this approach provides an effective and operationally viable framework for the cohesive post-processing of weather forecasts across multiple models and lead times to produce a well-calibrated probabilistic output.
\end{abstract}
\maketitle

\section{Introduction}

The importance of weather forecasting for decision support is likely to increase as we progress into times of changing climate and perhaps more frequent extreme conditions \citep{rahmstorf_increase_2011}. Any methodological developments that can improve our ability to make the optimal decisions in the face of meteorological uncertainty are likely to have a real impact on all areas that utilise weather forecasts.

Since the inception of meteorology as a mathematical science, driven by the likes of \citet{abbe_physical_1901}, \citet{bjerknes_problem_1904}, and \citet{richardson_weather_1922}, numerical modelling has been the core methodology of weather forecasting. In 2015, \citet{bauer_quiet_2015} reviewed the progress of numerical forecasting methods in \textit{the quiet revolution of numerical weather prediction}, and explained how improvements in physical process representation, model initialisation, and ensemble forecasting have resulted in average forecast skill improvements equivalent to one day's worth per decade --- implying that in 2020 our five day forecasts have approximately the same skill as the one day forecasts of 1980. 

However, the continuation of these gains requires ever more computational resources. For example, in pursuit of higher resolution models, halving grid cell length in three dimensions requires eight times the processing power, but due to model biases and initial condition uncertainty, corresponding improvements in forecasting skill are not guaranteed. At the same time, as society progresses we are placing greater emphasis on efficiency and safety in everything we do. In order for businesses to operate efficiently and in order to keep the public safe from meteorological hazards, there should be great emphasis on improving the functionality of weather forecasts as decision support tools --- and that means bridging the gap between deterministic NWP model outputs (including sparse ensembles from these) and fully probabilistic forecasting approaches suitable for supporting decision making through the use of decision theory \cite{economou_use_2016, simpson_decision_2016}. In essence, statistical approaches are key to optimal, transparent, and consistent decision making.

At the same time, while numerical weather prediction methodology has evolved gradually over the last century (hence \textit{`the quiet revolution'}), the last decade has seen significant developments in machine learning and its rise into the scientific limelight, with promising results being demonstrated in a wide range of applications \cite[e.g.][]{gulshan_development_2016, silver_mastering_2017, hey_machine_2020}. The catalyst for this new wave of machine learning can perhaps be attributed to the results of \citet{krizhevsky_imagenet_2012} in the Large Scale Visual Recognition Challenge (ILSVRC) of 2012, who demonstrated for the first time that deep neural networks --- with their ability to automatically learn predictive features in order to maximise an objective function --- could outperform existing state-of-the-art image classifiers based on hand-crafted features, which had been the established approach for previous decades. The parallels between the hand-crafted features in image classification, and the human choices that are made in all kinds of data processing pipelines --- including weather forecasting --- have inspired exploration into new applications of machine learning. In meteorology, could these tools relieve pressure from current model development and data processing bottlenecks and deliver a step-change in the rate of progress in forecasting skill?

Initial efforts using machine learning in the context of post-processing NWP model output have shown promising results \citep[e.g.][]{rasp_neural_2018, chapman_improving_2019, taillardat_calibrated_2016} in both probabilistic and deterministic settings. We believe that the greatest value of machine learning in weather forecasting lies in the probabilistic capabilities of these methods: not only do they have the potential to learn to improve forecasting skill empirically, but also to bridge the gap between traditionally deterministic forecasting approaches (i.e. numerical weather prediction) and the probabilistic requirements of robust decision support tools.

To this end, in this paper we demonstrate our framework for probabilistic weather forecast post-processing using machine learning. We have designed this framework to be suitable for use by operational meteorologists, and therefore, unlike other studies that we are currently aware of, our proposed solution incorporates forecast data from all available model solutions (i.e. multiple NWP model types, and all available forecast lead times). The framework aggregates the available forecast information into a single well-calibrated predictive distribution, providing probabilities of weather outcomes for each hour into the future. Our application is road surface temperature forecasting --- a univariate output --- using archived operational data from the UK Met Office. In this demonstration we use Quantile Regression Forests \citep[QRF, ][]{meinshausen_quantile_2006} as our machine learning algorithm, but hope to convince readers that our overall approach --- flexible quantile regression for each forecast, followed by averaging of quantiles across forecasts, and finally interpolating the full predictive distribution --- provides a flexible framework for probabilistic weather forecasting, and crucially one that is compatible with the use of any probabilistic forecasting models (post-processed or otherwise).

\begin{figure}[H]
	\includegraphics[width=1\textwidth]{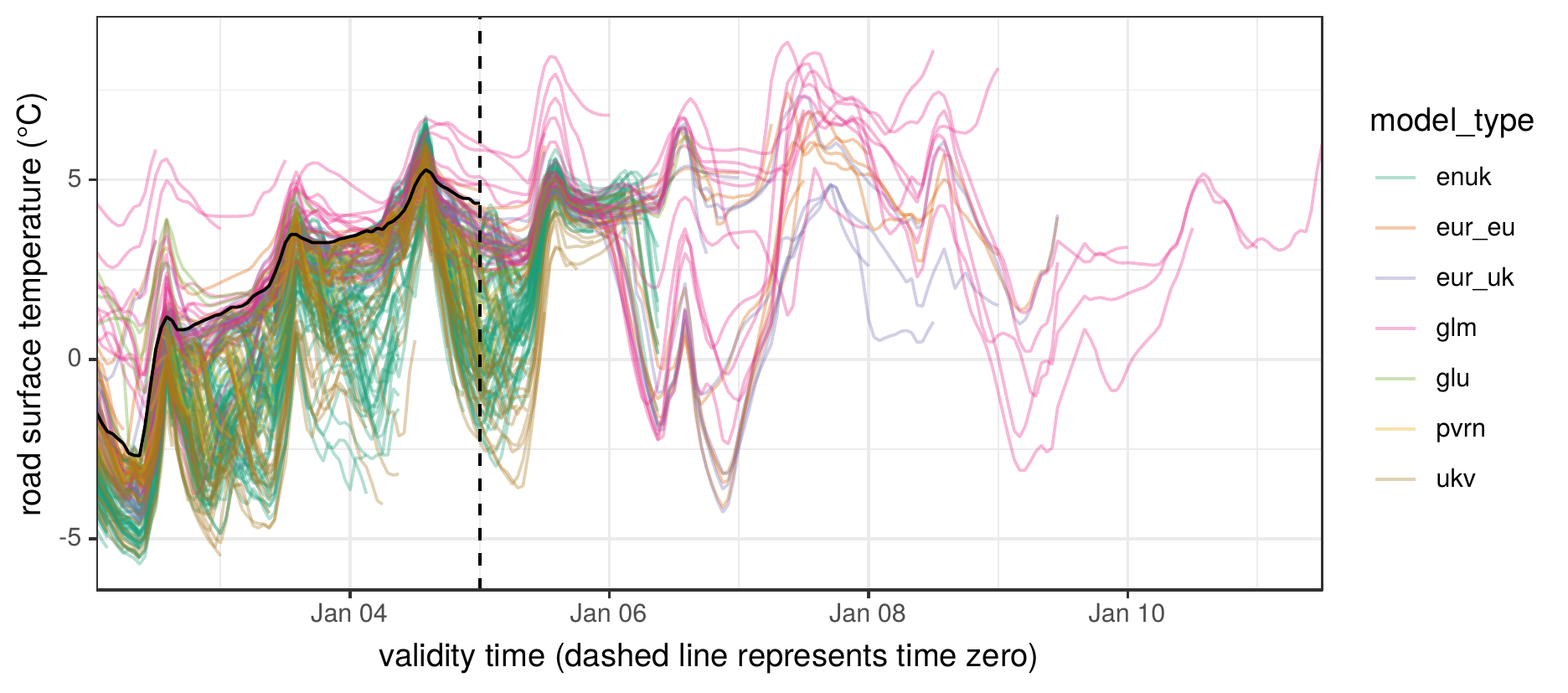}
	\caption{A visualisation of the information provided by numerical weather prediction (NWP) forecasts. Each coloured line represents an ensemble member from a different model type. Observations (solid black line) go as far as time zero (vertical dashed line - the `current time', which is 00:00 on 5th of Jan in this figure) and beyond that, if a statistical approach is not used, it's down to individual meteorologists to determine the likely weather outcomes based on the information presented by the models.}
	\label{thechallenge}
\end{figure}

Our framework can be seen as an overarching aggregator of forecast information, emulating part of the role of the operational meteorologist, who must otherwise develop a sense for how skillful each individual forecast is through experience, and mentally combine these forecasts in order to make probabilistic statements to inform decision making. These include judgements of uncertainty such as a `most likely scenario' and a `reasonable worst case scenario' \citep{stephens_improving_2014}. \autoref{thechallenge} gives an example of how complex a task it is to make sense of the available forecast information, even for the single variable of road surface temperature at a single site.

While methods for weather forecast post-processing using more traditional statistical approaches have existed for some time \citep[e.g.][]{raftery_using_2005, glahn1972use, wilks_comparison_2007, gneiting2005calibrated}, we believe our machine learning based approach to be a useful contribution to the field as interest in meteorological machine learning grows. The development of our framework has been guided by the needs of operational weather forecasting, including handling sets of different weather forecasting models with their own unique ranges of lead times. Increasingly these forecasts may not all be raw NWP forecasts, but are themselves likely to have been individually post-processed using machine learning (e.g. for downscaling), or purely statistical spatio-temporal forecasts. It is therefore a strength of our proposed framework that we can post-process any number of models of any type, and for any lead times.

\section{Post-processing framework}

The key considerations in designing our framework were that we wanted to develop an approach that was flexible, compatible, and fast. Flexible in the sense that we would like to minimise the number of assumptions made that would constrain the form of our probabilistic forecasts, and largely `let the data do the talking', as tends to be the machine learning ethos. Compatible in the sense that we would like our framework to generalise to scenarios in which NWP model outputs are not the only forecast available - this is likely to become more common as machine learning becomes more commonplace. And fast, because weather forecasting is a near-real-time activity and any post-processing approach has to be able to keep up.

There are many possible approaches for post processing individual weather forecasts, and indeed many possible approaches for producing forecasts in the first place (for example spatio-temporal statistical models \citep{hengl_spatio-temporal_2012}, or more recently neural network based approaches \citep{asanjan_short-term_2018}, in addition to the traditional NWP models). By using quantiles as the basis on which we combine multiple forecasts, our approach is compatible with any forecast from which well-calibrated predictive quantiles can be obtained, either from the forecast model directly (if  probabilistic), or through uncertainty quantification of deterministic models, as we demonstrate in this paper. The three stages of our framework's methodology are explained in the following subsections.

\subsection{From deterministic to probabilistic forecasts}

For our application to road surface temperature forecasting, the available forecasts come from a set of NWP models, as is commonly the case. Our model set spans from long range, low resolution global models (glu, glm) through medium range, medium resolution European models (eur\_eu, eur\_uk) to shorter range, high resolution UK specific models (ukv, enuk) including a six-hour nowcast (pvrn). Apart from the `enuk' model, which itself provides an ensemble of 12 members on each run, the other models provide single deterministic forecasts. While all of these models provide spatial forecasts, in this study we post-process the forecasts for specific sites in order to focus on the probabilistic aspects. \autoref{thechallenge} shows a snapshot of the set of model forecasts for a single site.

While the final output of our framework is a full predictive distribution summarising the information contained in the entire set of NWP model output, the first step is to convert each deterministic forecast into an individually well-calibrated probabilistic forecast. We do this by using machine learning to model the error profile of each deterministic forecast conditional on forecasting covariates. The error is defined as:
\begin{equation}\label{eq:error}
\epsilon_{t,m} = y - x_{t,m}
\end{equation}
where $x_{t,m}$ is a NWP model forecast for model type $m$ (e.g. `eur\_uk') and lead time $t$ while $y$ is the corresponding observation. For our surface temperature data, lead times range from 0 hours to 168 hours. Predictions of future data points are then obtained by
\begin{equation}\label{eq:prob}
\hat{y}_{t,m} = x_{t,m} + \epsilon_{t,m}
\end{equation}

\begin{figure}[H]
	\includegraphics[width=1\textwidth]{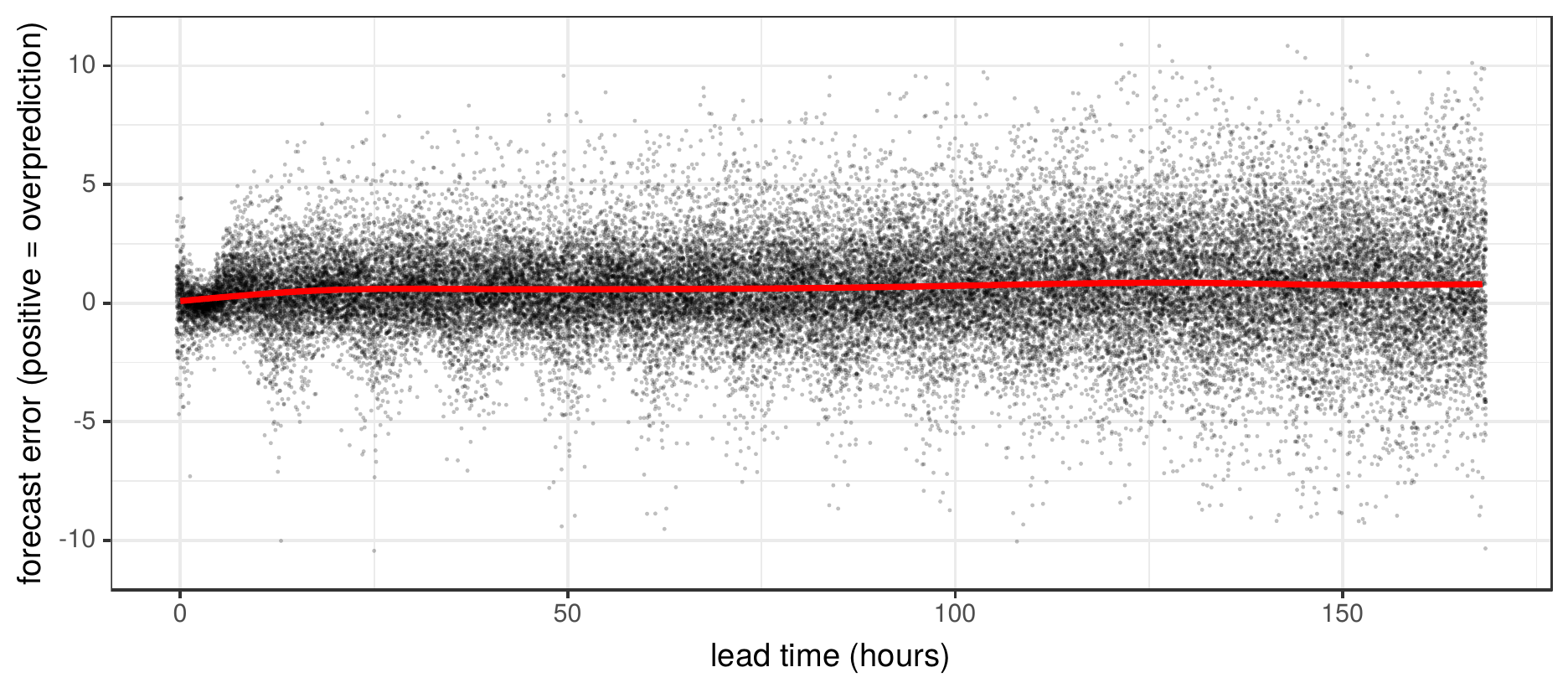}
	\caption{Plot of $\epsilon_{t,m}$ for $m=\mbox{glm}$ against lead hour $(1,2,\ldots,168)$ for a random sample of our dataset (spanning multiple months of absolute time). Each point is $\epsilon_{t,m}$ at a single hourly time step. The red line is a smooth estimate of the mean.}
	\label{leaderror}
\end{figure}

Modelling the forecast errors rather than $y$ was empirically found to produce better predictions using significantly less training data. An explanation for this is that $x_{t,m}$ is used as a complex trend removal function (e.g. for seasonality and other non-stationary effects), thus allowing us to treat $\epsilon_{t,m}$ as a time-invariant (stationary) variable --- the stochastic relationship between model error and lead time is quite stable across absolute time (see \autoref{leaderror}). This simplifying assumption may not hold up in every case, and we would recommend checks before applying it to other variables and forecasting tasks. Modelling the forecast errors, $\epsilon$, also has the benefit of providing many more unique $\epsilon_{t,m}$ observations for training than is provided by the absolute temperature observations $y_{t,m}$. This is because, while $y_{t}$ is identical for all $m$ (only one absolute temperature observation is made per time step), $\epsilon$ is unique for each $t,m$ pair because each unique NWP forecast produces its own unique error. The recent work of \citet{taillardat_research_2020}, and \citet{dabernig2017spatial} before them, shows that we are not alone in successfully using an error modelling approach.

Figure \ref{leaderror} shows $\epsilon_{t,m}$ for $m=\mbox{glm}$ (global long range forecast) and $t=0,1,\ldots,168$. Note the expected general increase in variance with increasing lead times and the increase in the location of the mean of the distribution (red line) indicating a systematic bias in the forecast. There is also a cyclic trend caused by the interaction between lead time and model initialisation time. This particular model is initialised at 00:00 and 12:00 hours, so we see increased errors on a 12 hour cycle starting from initialisation. This is because temperature errors tend to be larger in the early hours of the afternoon (when effects of inaccurately modelled cloud coverage on solar irradiance are most pronounced) compared to the early evening and morning.

In order to learn the error distribution of each NWP model type, we use Quantile Regression Forests \citep[QRF, ][]{meinshausen_quantile_2006} as implemented in the `ranger' package in R \citep{wright_ranger_2017}. While many other data modelling options are possible, QRF has a number of desirable properties. First, it has the flexibility to fit complex functions with minimal assumptions. For data rich problems such as ours, not specifying a parametric distribution allows us to capture the true complexity of the error distribution. Second, it is very fast in both training and prediction, and suitable for operational settings avoiding user input such as convergence checks (e.g. MCMC or gradient descent based methods). Third, it is relatively easy to understand the algorithm and has only a few hyper-parameters to tune, which makes getting reasonably good results in new problems quite straightforward.

For a detailed explanation of the QRF algorithm see \citet{athey_generalized_2019} or \citet{taillardat_calibrated_2016} for a more weather oriented description. For regression problems like ours, the QRF algorithm (a variant of the popular random forest algorithm) consists of an ensemble of regression trees. A regression tree recursively partitions the space defined by the covariates into progressively smaller non-overlapping regions. A prediction is then some property/statistic of the observations contained within the relevant region. Conventionally for each tree the prediction is the sample mean of the observations in the partition corresponding to new input data. Suppose for instance that a regression tree is grown on the data in \autoref{leaderror} and that our aim is to predict the mean forecast error at 100 hours. Suppose also that the tree had decided to group all observations in $t\in[98,106]$ into the same partition. Then the prediction for $t=100$ would simply be the mean of all observations between 98 and 106 hours. For a QRF however, the same tree would instead return the values of all the observations between 98 and 106 hours as an empirical distribution from which quantiles are later derived. The predictive performance of random forests is sensitive to how the covariate space is partitioned. The splitting rule, which governs the placement of partitioning splits as each tree grows, is therefore an important parameter, as are tunable hyper-parameters that we discuss in the next paragraph. Here we use the variance splitting rule, which minimises the intra-partition variance within the two child partitions at each split. A key aspect of the random forest and QRF algorithm is that each tree in the ensemble is grown on its own unique bootstrapped random sample of the training data. This produces a forest of uncorrelated trees, which when aggregated (called bootstrap aggregation or `bagging') results in an overall prediction that is less prone to over-fitting than an individual decision tree, while retaining the ability to learn complex functions. To produce quantile predictions, the QRF returns sample quantiles from all observations contained within the relevant partition of each individual tree in the forest. In doing so it behaves as a conditional (on the covariates) estimate of the CDF.

\begin{figure}[H]
	\centering
	\includegraphics[width=1\textwidth]{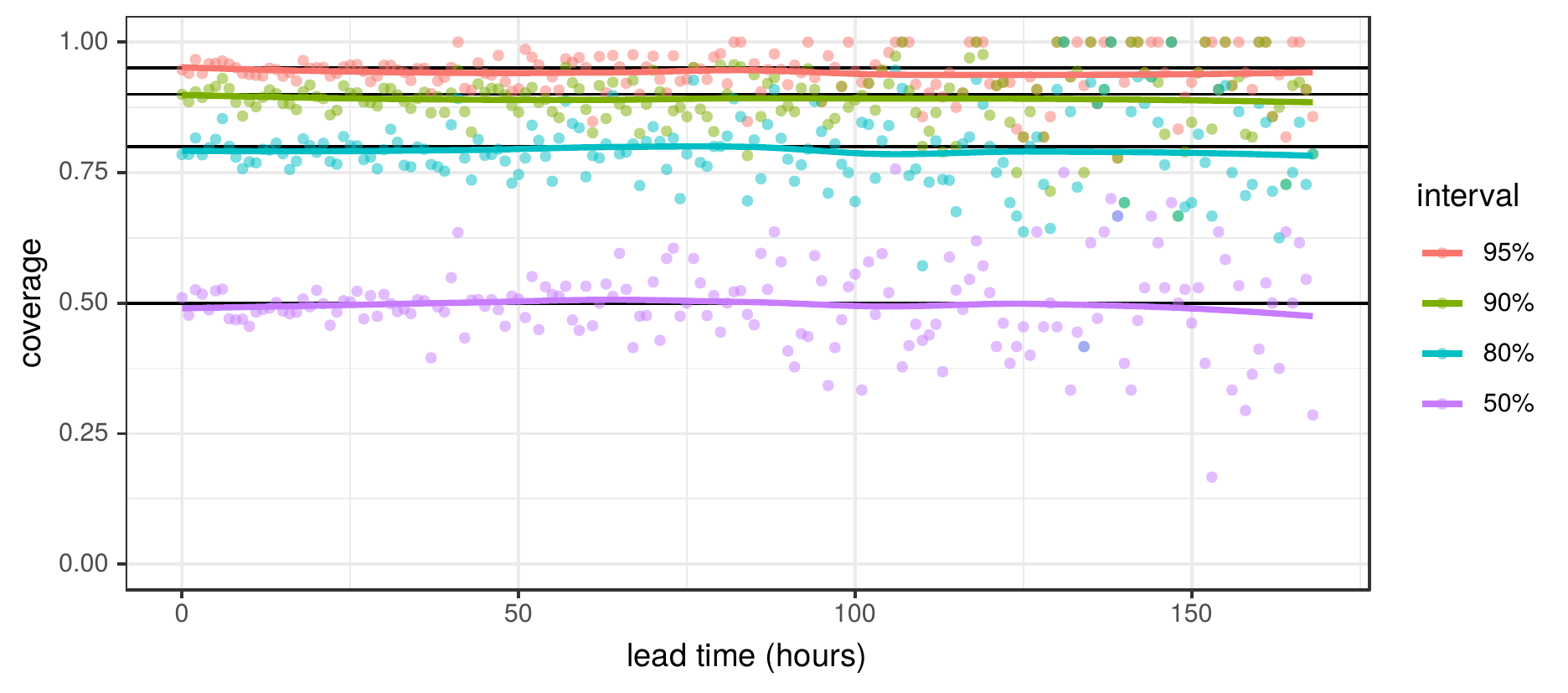}
	\caption{Coverage of the 50\%, 80\%, 90\%, and 95\% QRF prediction intervals on out-of-bag data from one training scenario (though the picture is indicative of other scenarios). The coverage is the proportion of observations that fall within each prediction interval, and should match the interval (i.e. with 95\% of observations falling within the 95\% prediction interval) in a well-calibrated setup.}
	\label{oobcov}
\end{figure}

For modelling NWP surface temperature errors, the tuning of QRF hyper-parameters as well as the selection of input covariates was conducted manually with the aim of achieving good out-of-bag quantile coverage (a QRF proxy for out-of-sample performance) across all lead times. This was achieved using visual checks such as \autoref{oobcov}, which indicates that on average, prediction intervals are close to the ideal coverage across lead times, i.e. 90\% of the time observations will fall within the 90\% prediction interval. However for operational setups it may be preferable to use a more formal optimisation procedure, such as Bayesian optimisation. We found that using just lead time, $t$, and model type, $m$, as covariates gave the best calibration results, presumably aided by the parsimonious nature of this simple representation. The chosen hyper-parameters were: mtry = 1 (this is the number of covariates made available at random to try at each split), min.node.size = 1 (this limits the size of the terminal nodes / final partitions of each tree - in this case there is no limit on how small these can be), sample.fraction = 128/nrow(training data) (this is the size of the bootstrap sample of the training data provided to each tree), and num.trees = 250 (this is the number of trees in the forest). The use of a relatively small sample size (128 observations for each tree, out of a total of around 50,000 observations in a 14 day run-in period) and a minimum node size of one (trees grown to full depth) was found to produce the best out-of-bag coverage at a minimal run time. Our mtry setting meant that one of our two covariates ($t$ and $m$) was made available at random to each tree at each split. If another objective had been prioritised (e.g. to minimise mean squared error, rather than optimise coverage) the optimal hyper-parameters would be different.

\begin{figure}[H]
	\includegraphics[width=\textwidth]{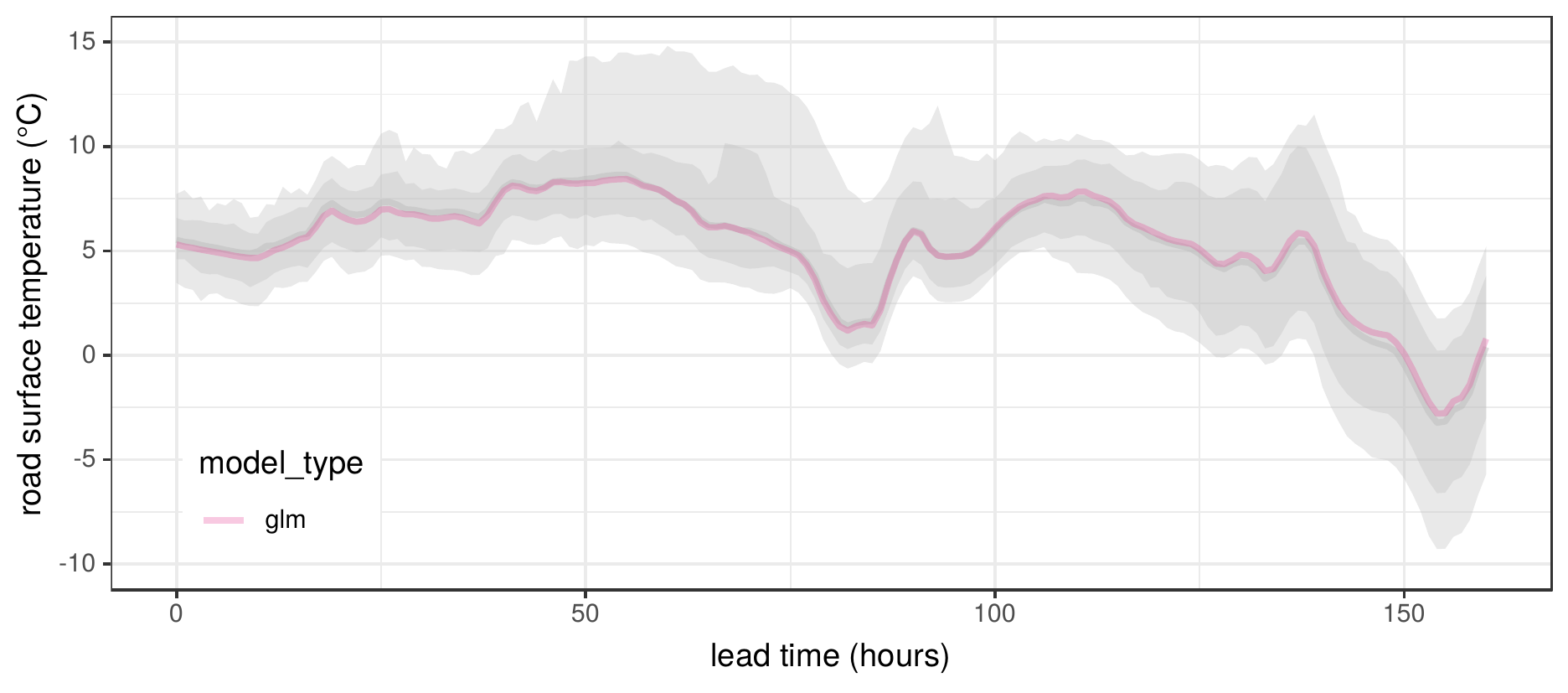}
	\caption{A deterministic NWP forecast for $m=\mbox{glm}$ that has been converted to a probabilistic forecast using equation \eqref{eq:prob}. The 80\% and 95\% prediction intervals are shown as overlain grey ribbons, while the solid grey line is the median (which differs little from the NWP forecast here).}
	\label{conversion}
\end{figure}

Once the QRF has been trained, each NWP forecast can be converted to a probabilistic forecast by adding to it the predicted error distribution \eqref{eq:prob}. Unlike the deterministic NWP forecast, the prediction is now a probability distribution, constructed through a conditional bootstrap of $\epsilon_{t,m}$ via the QRF algorithm. Prediction intervals are obtained as quantiles of this distribution as illustrated in \autoref{conversion}.

\subsection{Combining probabilistic forecasts}

The next step is to combine these predictive distributions from each NWP model output into a single distribution that is suitable for use in decision support. The challenge is to combine the forecasts in a probabilistically coherent manner, with the goal of producing a single well-calibrated and skillful predictive distribution.

\begin{figure}[!htb]
	\centering
	\includegraphics[width=1\textwidth]{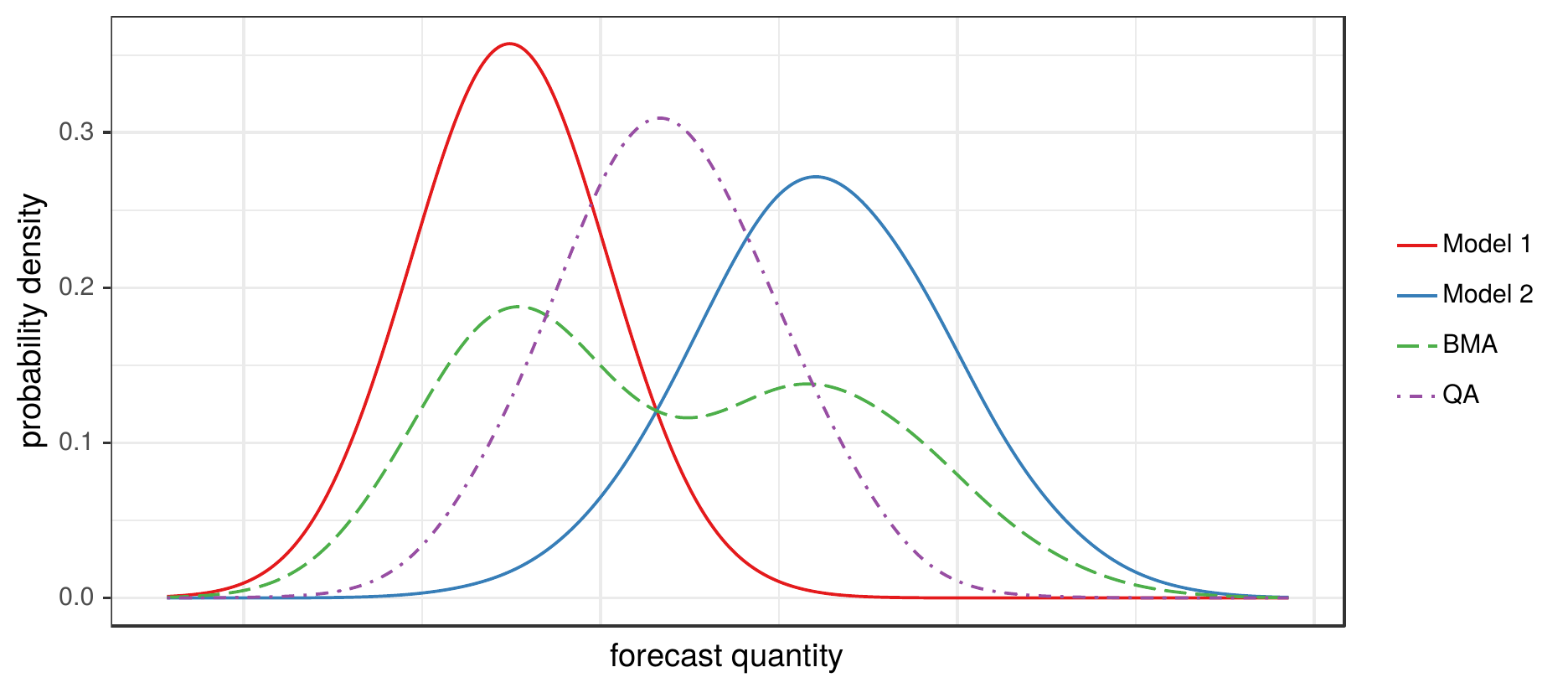}
	\caption{Synthetic example of combining two probabilistic forecasts using Bayesian Model Averaging (BMA) and Quantile Averaging (QA), after \cite{schepen_model_2015}.}
	\label{bmaqma}
\end{figure}

A popular approach for combining probabilistic models is Bayesian Model Averaging (BMA), and its use in the statistical  post-processing of weather forecasts has precedent \citep[e.g.][]{hoeting_bayesian_1999, raftery_using_2005, gneiting_combining_2013}. Basic BMA produces a combined distribution as a weighted sum of PDFs. However, in order to satisfy the requirements of our framework, we propose an alternative approach using quantile averaging, whereby each quantile of the combined distribution is taken as the mean of the same quantile estimated by each individual model. An illustrative comparison of equal-weighted BMA and quantile averaging is shown in (\autoref{bmaqma}). For the purposes of our framework, we found BMA to be unsuitable for the following three reasons: 1) Achieving good calibration of the combined distribution produced by BMA requires optimisation of the intra-model variance, i.e. the spread of each individual model's error profile. In our case, where each model's error profile has been learned independently by QRF, and is already well-calibrated, combining these through BMA produces an over-dispersed predictive distribution due to the inclusion of the inter-model variance in addition to the already calibrated intra-model variances. 2) In turn, this makes BMA rather incompatible with input models that are individually well-calibrated (e.g. statistical nowcasts), and therefore incompatible with a general framework like ours. 3) The use of BMA across all models and lead times is complicated by the fact that there are not an equal number of forecasts available for each lead time. This means that the inter-model variance is intrinsically inconsistent across lead times, even dropping to zero at our longest ranges, where only a single deterministic forecast is available (e.g. \autoref{thechallenge}). This decrease in inter-model variance with increasing forecast range trends opposite to the true uncertainty, which intuitively should increase with forecast range. This is a quirk of NWP forecast availability and one that probabilistic post-processing must overcome.

Our framework overcomes this instability in inter-model variance by using quantile averaging (also known as the `Vincentization' method \citep{genest_vincentization_1992, vincent_function_1912}) to combine forecasts that are already well-calibrated for coverage (owing to their QRF error profiles, in our case). Using this approach, we construct our combined forecast distribution from the quantile predictions of our individual QRF post-processed forecasts. To produce each predicted quantile of the combined distribution, Vincentization simply takes the mean of the set of estimates of the same quantile by each individual forecast. As explored by \citet{ratcliff_group_1979}, Vincentization produces a combined distribution with mean, variance, and shape all approximately equal to the average mean, variance, and shape of the individual distributions (as we see in \autoref{bmaqma}). Vincentization therefore provides similar functionality to parameter averaging of parametric distributions, but for non-parametric distributions such as ours. Within our framework, Vincentization effectively integrates out the inter-model variance (by taking the mean across models), and in doing so preserves the calibration of the individual QRF post-processed forecasts, avoiding the overinflation issues that BMA would produce. Vincentization is therefore one possible solution to the issue of combining calibrated probability distributions without loss of calibration \cite{gneiting_combining_2013}. However, the method by which probability distributions are combined can have important implications for decision-support forecasting, and while quantile averaging satisfies our general requirements for this framework, we do not discount that alternative approaches may be preferable depending on the application.

Our quantile averaged forecast benefits from stability owing to the law of large numbers --- any quantile of the forecast distribution represents an average of the estimates of that quantile across the available individual forecasts. This approach is therefore more akin to model stacking procedures, as used in ensemble machine learning to improve prediction accuracy by reducing prediction variance \citep{ren_ensemble_2016}. Indeed, this same logic is behind the bootstrap aggregation (`bagging') procedure of the random forest algorithm: by averaging the predictions of multiple individual predictors --- each providing a different perspective on the same problem --- the variance of the aggregate prediction is reduced, resulting in improved prediction accuracy at the expense of some increased bias \citep{belkin_reconciling_2019}. Crucially for our framework, unlike a BMA approach which retains the inter-model variance, the calibration of our quantile averaged output is invariant to the number of forecasts available at each timestep. This is key for temporally coherent forecast calibration across all lead times.

Our error modelling approach does require one extra-step of processing in order to handle model types which themselves have multiple interchangeable ensemble members. The `enuk' model (\autoref{thechallenge}) is our example of this, having twelve non-unique members. In such cases, the apparent error profile for the model type as a collective gets overinflated by the inter-member variance. Our solution to this is to label each ensemble member by its rank (at each time step). This splits our 12-member `enuk' ensemble into 12 unique model types in the eyes of the QRF. This approach produces well-calibrated error profiles (though with significant offset bias in the extreme ranking members, as would be expected).

\subsection{Simulation from the full predictive distribution}

\begin{figure}[!htb]
	\includegraphics[width=0.5\textwidth]{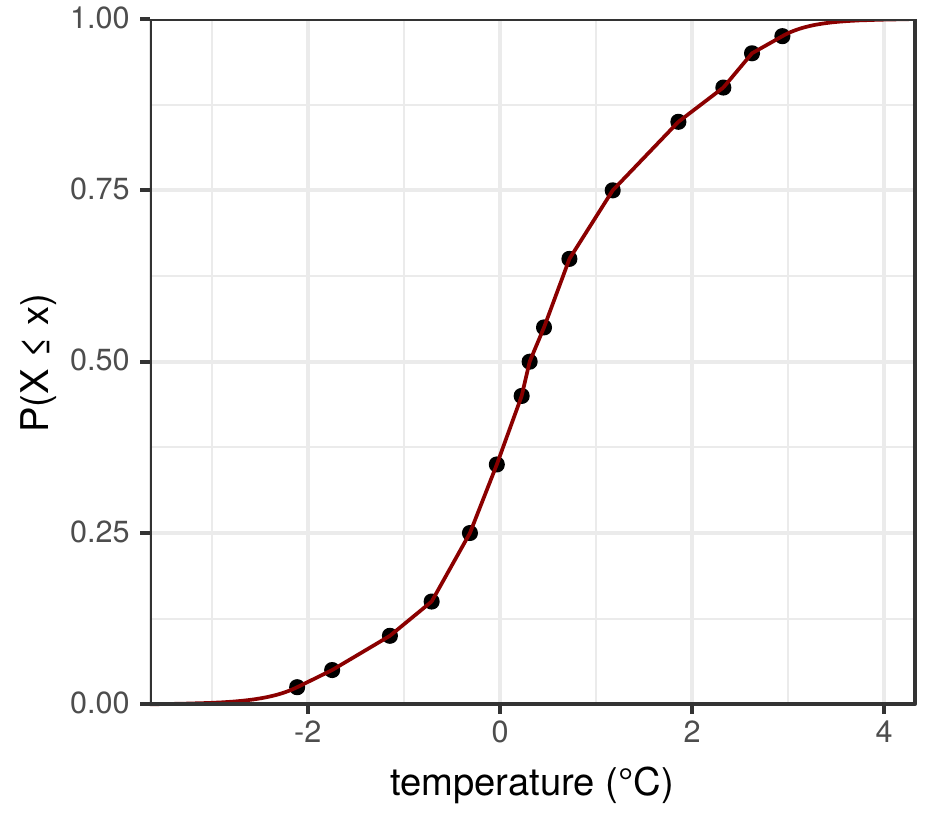}
	\includegraphics[width=0.5\textwidth]{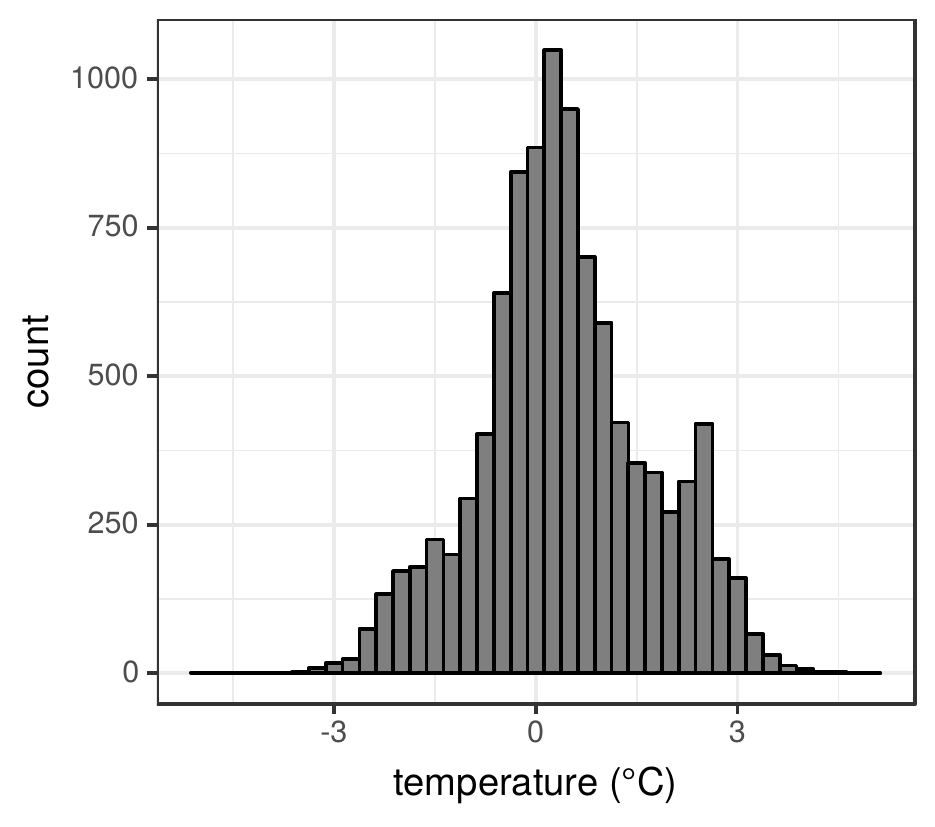}
	\caption{Interpolated CDF of the combined predictive distribution (left), and corresponding road surface temperature simulation (right) for a particular 50-hour ahead forecast.}
	\label{cdfinterp}
\end{figure}

While quantile averaging provides an effective way of combining multiple probabilistic forecast distributions, it leaves us with only a set of quantiles rather than the full predictive distribution. This distribution is desirable because it allows us to (a) answer important questions such as `what is the probability that the temperature will be below $0^\circ$C?' and (b) evaluate the skill of the probabilistic forecast using a range of proper scoring rules (although, depending on the end use, some proper scoring rules could be calculated directly from quantile predictions, e.g. the quantile score \cite{bentzien2014decomposition} or the interval score \cite{gneiting_strictly_2007}). 

To obtain the full predictive distribution, we interpolate between the quantiles of our combined forecast in order to construct a full CDF using the method of \citet{quinonero-candela_evaluating_2006}, which has previously been applied to precipitation forecasting \cite{cannon_quantile_2011} and is available in the R package qrnn \cite{cannon_quantile_2011}. The method linearly interpolates between the given quantiles of the CDF (our combined quantiles from Vincentization), and, beyond the range of given quantiles, extrapolates down to $P(X \leq x) = 0$ and up to $P(X \leq x) = 1$ assuming tails that decay exponentially with a rate that ensures the corresponding PDF sums to one (\autoref{cdfinterp} left, for details see pages 8 and 9 of \citet{quinonero-candela_evaluating_2006}). Using this approach allows us to construct a full predictive distribution from the Vincentized quantiles of our individual QRF post-processed forecasts. Depending on the application at hand, suitable forecast information might be obtained by querying the CDF of the predictive distribution directly at each time step, but in our application here, we go the extra step of simulating temperature outcomes at each timestep by randomly sampling from the CDF (\autoref{cdfinterp} right). This is the final step of our framework --- taking us from a set of disparate NWP forecasts to a full predictive distribution of weather outcomes.

\section{Results}

To evaluate our framework, we applied it to 200 randomly time-sliced and site-specific forecasting scenarios extracted from our UK Met Office road surface temperature dataset, which we have aggregated to hourly time steps. Each scenario has its own training window of 14 days, providing approximately 50 000 data points of $\epsilon_{t,m}$ to train the QRF, immediately followed by its own evaluation window extending as far as the longest range NWP forecast (up to 168 hours / 7 days), which is akin to the area to the right of the vertical dashed line in \autoref{thechallenge}. While there are only 336 hours in a 14 day training window, the number of NWP models and their regular re-initialisation schedule, means that approximately 150 forecasts are made for any hour by the time it is observed. While we only use the current forecasts from each model type to generate our predictions, the training benefits from every historical forecast within the window.

\begin{figure}[!htb]
	\includegraphics[width=1\textwidth]{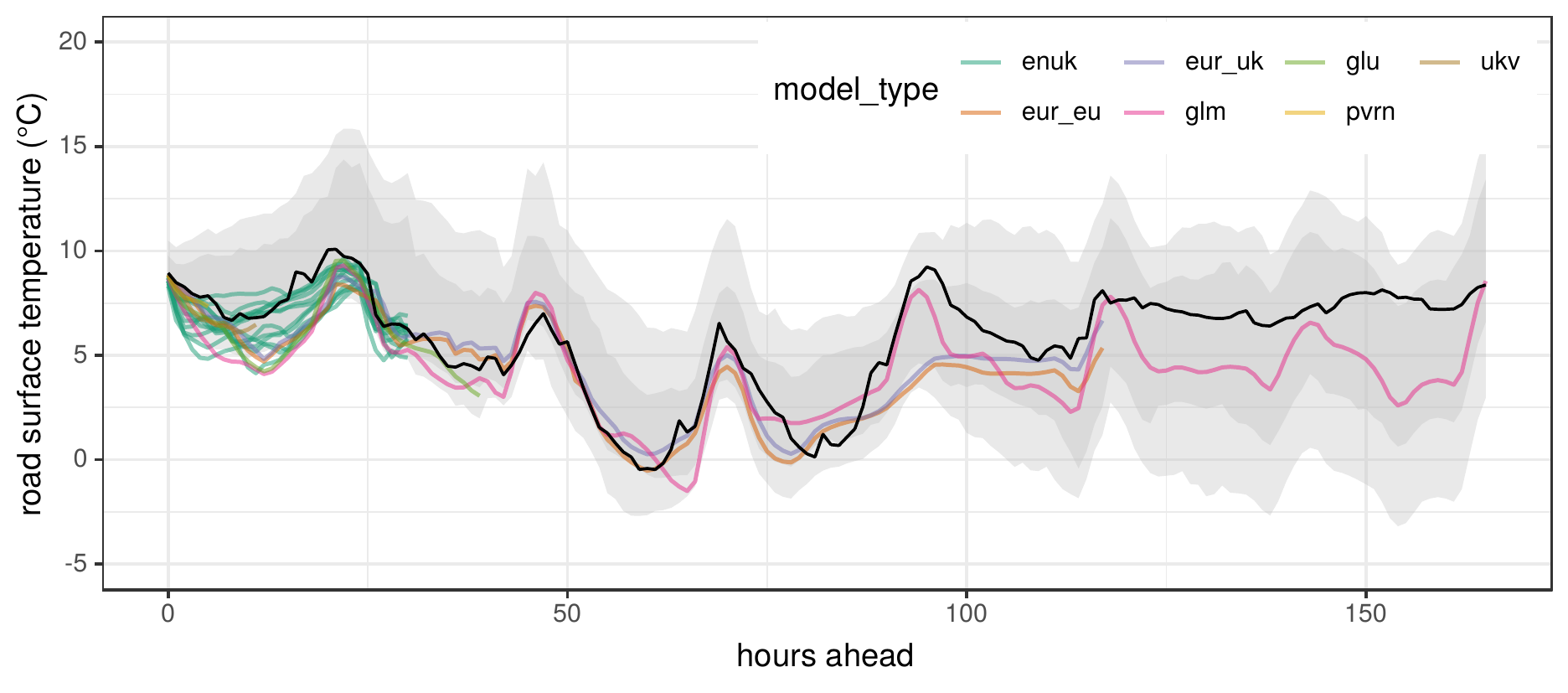}
	\includegraphics[width=1\textwidth]{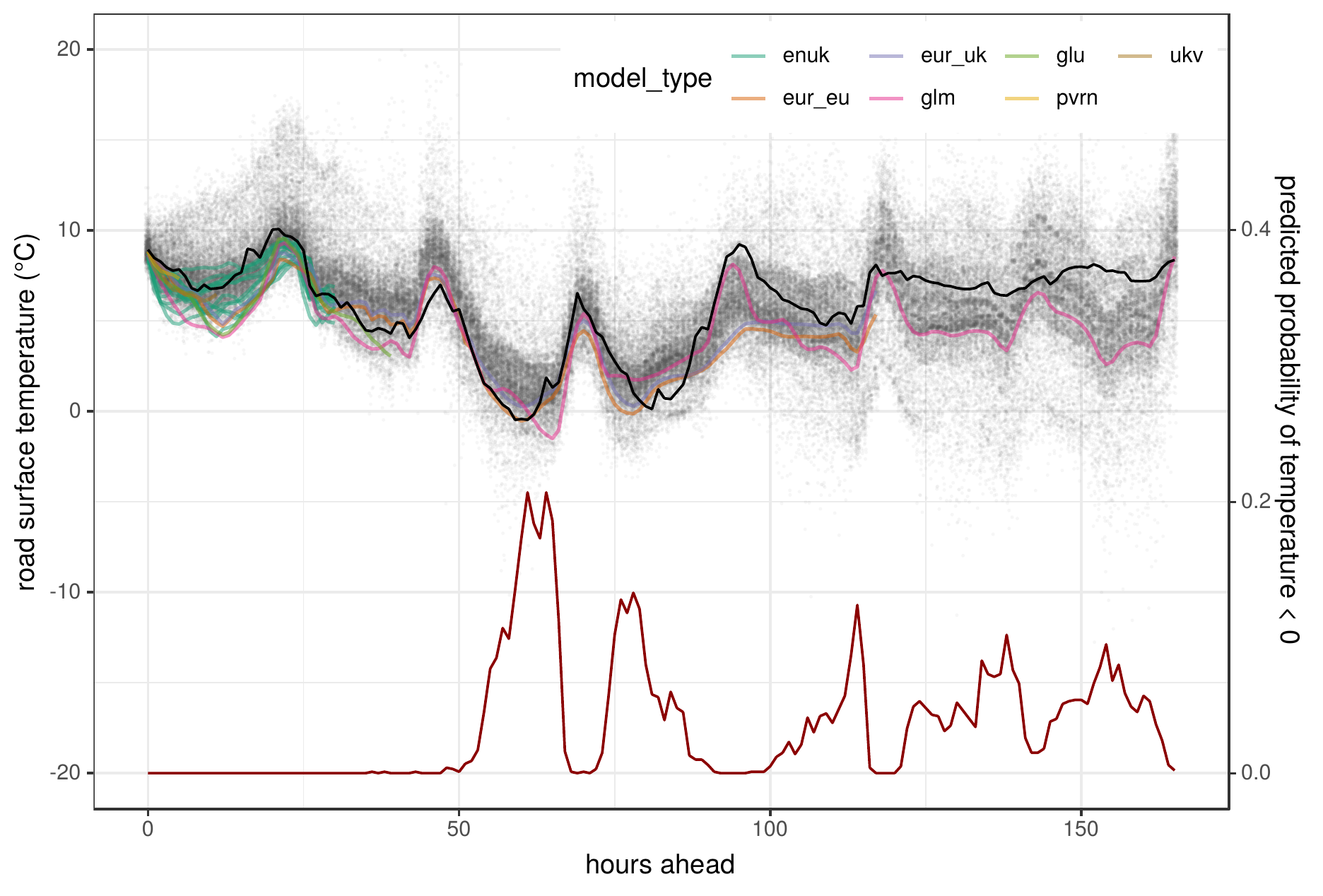}
	\caption{An example of the output of our post-processing framework. Top: the probabilistic forecast is visualised by the 80\% and 95\% prediction intervals. Bottom: simulations from the full predictive distribution as grey dots, while the red line (right-hand y-axis) shows the probability of temperature being $<0^\circ$C. NWP model forecasts are shown by coloured lines, and the true observed temperature (not known at time of forecasting) is shown by a solid black line.}
	\label{exampleforecast}
\end{figure}

\autoref{exampleforecast} shows an example prediction of up to 168 hours into the future for a particular scenario. This is just one of the 200 random scenarios used in our overall evaluation. Although the prediction at each hour ahead is a full probability distribution, here we present prediction intervals as well as a simulation of 1000 temperature values from it. The samples were used to derive the probability of the temperature being below $0^\circ$C as the proportion of values less than zero. Different stakeholders will require their own unique predictive quantities, and by providing a full predictive distribution, our framework should cater for a wide variety of requirements.

\begin{figure}[!htb]
	\includegraphics[width=0.5\textwidth]{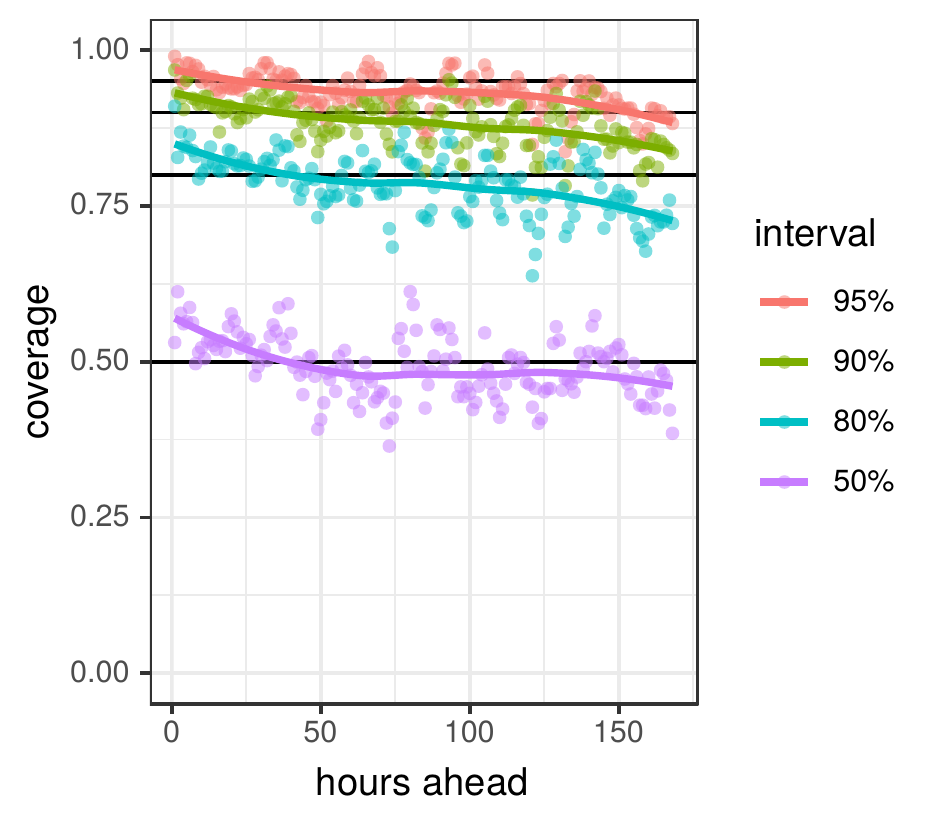}
	\includegraphics[width=0.5\textwidth]{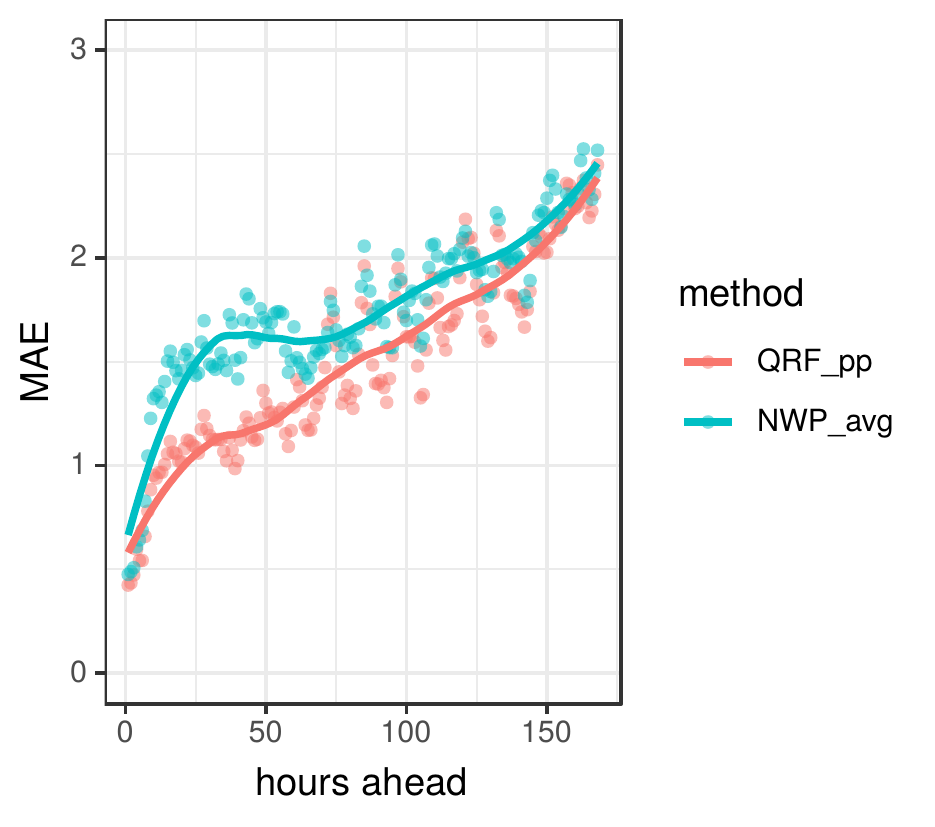}
	\caption{Evaluation metrics from 200 forecast scenarios. On the left, coverage of the prediction intervals of the combined probabilistic forecasts. On the right, the MAE achieved by the median of the combined probabilistic forecast (QRF\_pp) compared to taking the median of the available NWP forecasts (NWP\_avg).}
	\label{coverage}
\end{figure}

\begin{figure}[!htb]
	\includegraphics[width=1\textwidth]{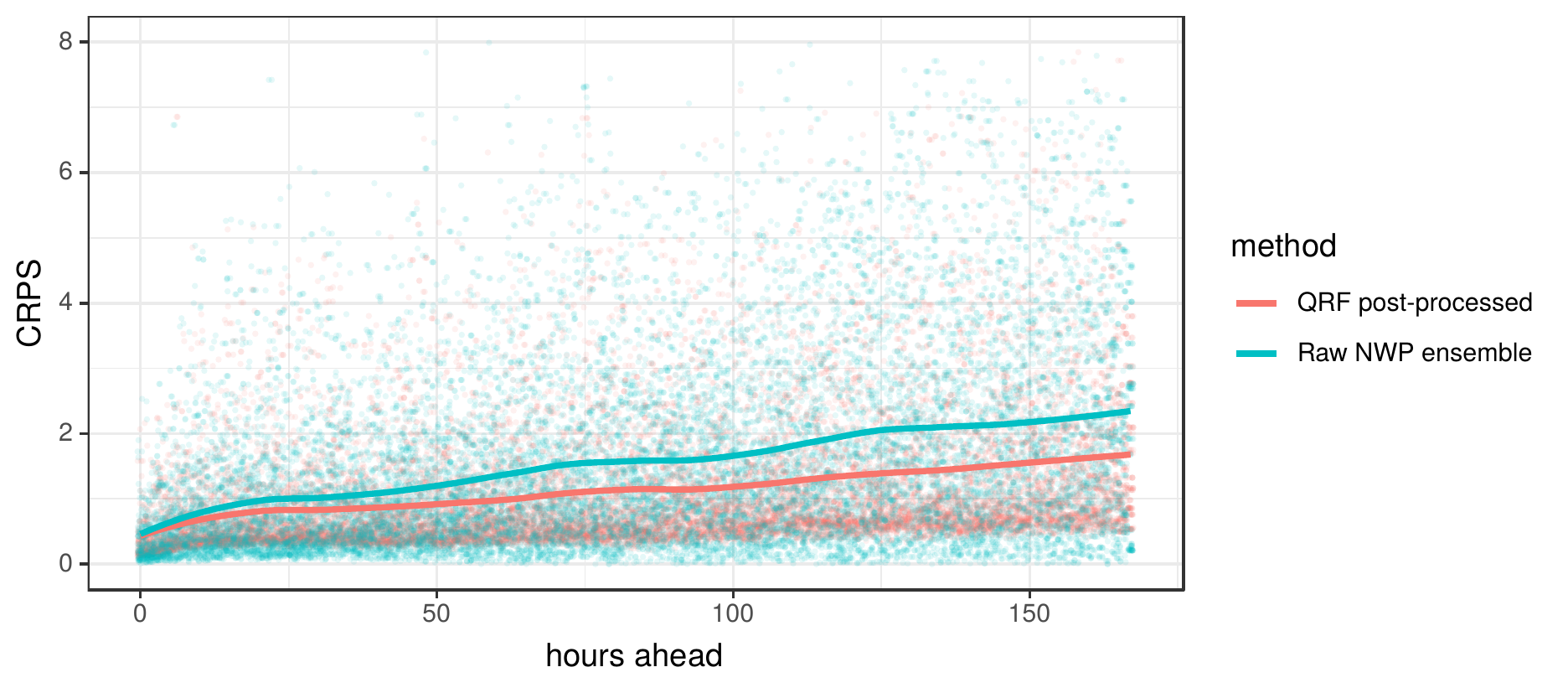}
	\includegraphics[width=1\textwidth]{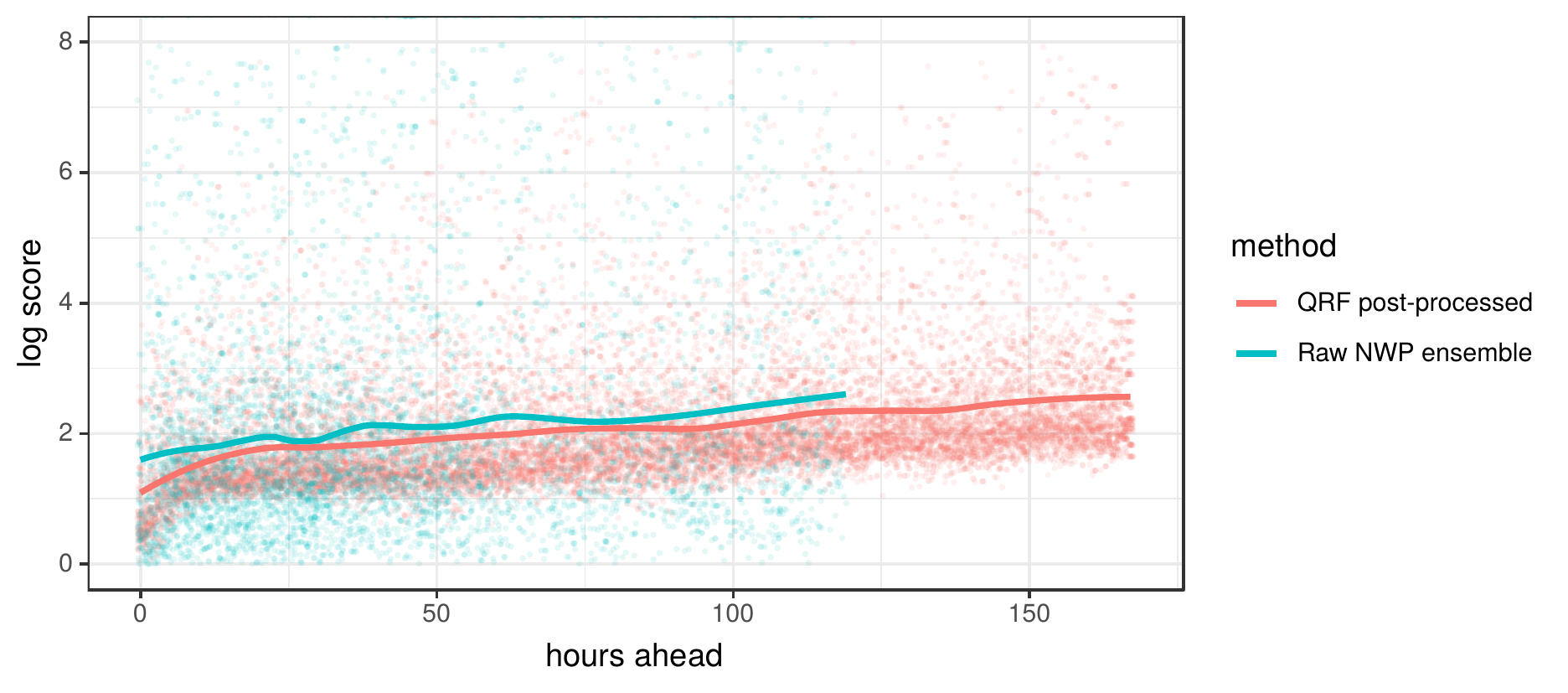}	
	\caption{Evaluation metrics of our post-processing framework across all lead times on 200 random forecast scenarios. We compare our QRF post-processed output to the raw NWP ensemble in terms of continuous ranked probability score (CRPS, top) and logarithmic score (bottom).}
	\label{CRPS}
\end{figure}

Various metrics could be used to evaluate the skill of our probabilistic forecasts over multiple scenario runs. From the perspective of decision support, the ideal metric to evaluate would be the change in loss resulting from using our forecasts to make real world decisions, such as about when to grit roads in our case. However, in the interest of a more general analysis we use a range of standard metrics. These are: prediction interval coverage (\autoref{coverage} left), the mean-absolute-error (MAE) of the median (\autoref{coverage} right, because sometimes a single `best' deterministic forecast is still desired), as well as the continuous ranked probability score (CRPS) and logarithmic score of our probabilistic forecast (both in \autoref{CRPS}).

\autoref{coverage} indicates that coverage is good overall, with 94.7\% of observations falling within the 95\% prediction interval, although there is some over-dispersion of our forecast at the shortest ranges and under-dispersion at the longest ranges. This is an indication that, despite producing near perfect results on out-of-bag training data (\autoref{oobcov}), the QRF performance diminishes slightly when applied to new data. The range dependent over- and under-dispersion may be due to the partitioning process on which the forest is grown - by necessity the partitions that represent the extremes of forecast range must extend some distance towards the middle of the range, and in doing so end up capturing an empirical error distribution that is slightly biased towards the average empirical error distribution, rather than perfectly representing the distribution at the extremes of covariates. It may be the case that other data modelling approaches could do better in this respect.

Although deterministic performance was not our focus, the QRF median prediction does outperform the median of the available NWP models across the entire forecast range in terms of MAE. While only a conceptual benchmark, this can be taken as some indication that we have not `thrown away' deterministic performance in pursuit of probabilistic calibration. \autoref{coverage} also indicates that our method results in a monotonically increasing error with forecast range, unlike the median of the original NWP forecasts. Similarly, we see a monotonic increase in both the CRPS and the logarithmic score with increasing forecast range (\autoref{CRPS}, top and bottom), and, when compared to the performance of the raw NWP ensemble on the same metrics, find our QRF post-processing approach to perform better. In the case of CRPS, our QRF post-processing approach reduces the rate at which forecasting skill decreases with forecast range. Also, by looking at the spread of performance across individual forecasting scenarios (represented by individual points in \autoref{CRPS}, rather than the lines, which trace the mean) we can see that our QRF post-processing approach reduces the variance in forecasting skill across different forecasting scenarios, making it a more consistent forecast than raw NWP. In the case of logarithmic score (\autoref{CRPS}, bottom) we see again that the forecasting skill provided by the QRF post-processed output is more consistent than that of the raw NWP ensemble, although the difference in the mean performance is less pronounced. The logarithmic score of the NWP ensemble cannot be obtained at longer ranges as only a single deterministic forecast is available. The authors recognise that comprehensive comparisons of our approach to other probabilistic post-processing approaches (in addition to raw NWP output) will be important to consider when choosing the best approach for any operational setup. While we do not offer such comparisons in this study, we have made our dataset openly accessible as one of several benchmark datasets compiled by \citet{haupt_towards_2020} at https://doi.org/10.6075/J08S4NDM in the hope that it will facilitate comparison of different post-processing approaches on common benchmarks in the future.

In terms of speed, training the QRF for each forecast scenario takes between just three and four seconds on an i7-8550U laptop, and so the implementation of this framework can be expected to add very little overhead to a typical operational NWP forecasting setup.

\section{Discussion and conclusions}

The conversion of disparate forecasts into a cohesive probabilistic output is important. A key function of weather forecasts is to support decision making, but current numerical methods do not provide the well-calibrated probabilistic output required to do this rigorously. By  applying our framework we compensate for this shortcoming, effectively supplementing forecasts with information from their historic performance in order to combine all available deterministic inputs, for all lead times, into a single well-calibrated probabilistic forecast. Whilst our approach is by no means the first to provide probabilistic post-processing of weather forecasts, we believe the flexibility and speed provided by our use of machine learning, along with our framework's relative simplicity and ability to simultaneously deal with all available models and lead times, makes it a strong option for consideration in operational forecasting settings.

In this study we have only applied our framework to site specific forecasting, but there are no fundamental reasons why the same principles cannot be applied to spatial forecasting by providing the QRF with additional spatial covariates against which to learn its error profiles, or by adopting the standardised anomaly model output statistics (SAMOS) approach as proposed by \citet{dabernig2017spatial}. The error modelling approach that we use seems a very effective way of minimising the amount of training data required compared to predicting absolute values. \citet{taillardat_calibrated_2016}, who also make use of QRF in their post-processing, initially used four years of training data for their absolute value forecasting system in 2016, but have since adopted an error modelling approach themselves \citep{taillardat_research_2020}.

There are still several aspects of our framework that are open to further investigation. One significant aspect that we explored in preliminary experiments but have not included in our methodology here, is the opportunity to use weighted quantile averaging for combining forecasts. In our setup, where all of the inputs are recent NWP forecasts (and therefore similarly skillful), we saw negligible difference in using a weighted averaging approach, but in situations where more diverse forecast types are in use, it may prove beneficial to assign weightings according to forecast skill. A dynamic weighting approach also enables individual models to be updated without jeopardising the overall post-processed output, as the contribution of the new or updated model will be minimal until it's error profile is well understood. The QRF algorithm provides a convenient means by which skill can be estimated ahead of time, in the form of out-of-bag metrics. For example, we showed earlier the out-of-bag coverage of our trained QRF (\autoref{oobcov}). Metrics such as the CRPS, logarithmic score, and Kullback–Leibler divergence would provide good comparisons of forecast skill on which to base quantile averaging weight, although their calculation would add some additional processing time. \citet{yao_using_2018} provide more detail about using such metrics for weighted model stacking, and in fact these weights can be optimised as an additional supervised learning problem \citep{ren_ensemble_2016}.

The overall strategy for combining forecasts is also open to further research. Because it retains the inter-model variance, BMA may be considered to provide a better representation of extreme outcomes at the expense of well-calibrated coverage (at least in setups where each input forecast is already well-calibrated, which is likely to become the norm). We also think that the output of BMA would be difficult to make use of in practice when applied across all lead times as in our framework, because of the discrepancy in the number of models available at each time step, and therefore the spurious inconsistency of the inter-model variance across the forecast range. Still, applications where capturing extremes is a priority may wish to investigate further. For general purposes, we are satisfied with our time-consistent and calibration-preserving quantile averaging approach.

It is our belief that, as time goes on, and the number of different forecasting models in use --- along with their complexity and resolution --- continues to increase, there will be increasing need for algorithmic interfaces such as ours to summarise the otherwise overwhelming sea of forecast information into decision ready output. This would consist of optimally well-calibrated probabilities of future weather outcomes given all available information. Probabilistic machine learning is a technology that can enable this, and we hope that the work we have demonstrated here will go some way in aiding progression towards this goal.\vskip6pt

\enlargethispage{20pt}

\dataccess{Our dataset has been made available with permission from the Met Office and Highways England, for which we are grateful. It is available for download along with several other open weather forecast post-processing datasets collated by \citet{haupt_towards_2020} at https://doi.org/10.6075/J08S4NDM. In addition, the code for this study can be accessed at https://github.com/charliekirkwood/mlpostprocessing}.

\aucontribute{All authors shared in conceiving and designing the study. CK wrote the code and manuscript, and ran the experiments behind the solution we present here, with other authors providing supervision. All Authors read and approved the manuscript.}

\competing{The authors declare that they have no competing interests.}

\funding{This work has been conducted as part of CK's PhD research, funded by the UK's Engineering and Physical Sciences Research Council (EPSRC project ref: 2071900) and the Met Office.}

\ack{The lead author is grateful for the insightful discussions and community feedback that came from attending the Machine Learning for Weather and Climate Modelling conference at the University of Oxford in 2019, and the EUMETNET Workshop on Artificial Intelligence for Weather and Climate at the Royal Meteorological Institute of Beligum in 2020. In addition CK would like to thank Thomas Voigt for offering his data engineering expertise. We are grateful to all who have helped to guide this study.}


\bibliography{PhD.bib}

\end{document}